\documentclass[sigconf]{acmart}
\usepackage{booktabs} % For formal tables
\usepackage{enumitem}
\usepackage{balance}
\usepackage[utf8]{inputenc}
\usepackage{multirow}
\usepackage{graphicx}
\usepackage{microtype}
\usepackage{subfigure}
\usepackage{bm}
\usepackage{amsmath}
\usepackage{amsfonts}
\usepackage{graphicx}

\newcommand{\squishlist}{
 \begin{list}{$\bullet$}
  { \setlength{\itemsep}{0pt}
     \setlength{\parsep}{3pt}
     \setlength{\topsep}{3pt}
     \setlength{\partopsep}{0pt}
     \setlength{\leftmargin}{1.5em}
     \setlength{\labelwidth}{1em}
     \setlength{\labelsep}{0.5em} } }

\newcommand{\squishlisttwo}{
 \begin{list}{$\bullet$}
  { \setlength{\itemsep}{0pt}
     \setlength{\parsep}{0pt}
    \setlength{\topsep}{0pt}
    \setlength{\partopsep}{0pt}
\setlength{\leftmargin}{2em}
\setlength{\labelwidth}{1.5em}
\setlength{\labelsep}{0.5em} } }

\newcommand{\squishend}{
\end{list}  }

\AtBeginDocument{%
  \providecommand\BibTeX{{%
    \normalfont B\kern-0.5em{\scshape i\kern-0.25em b}\kern-0.8em\TeX}}}

\setcopyright{acmcopyright}
\copyrightyear{2018}
\acmYear{2018}
\acmDOI{10.1145/1122445.1122456}

\acmConference[Woodstock '18]{Woodstock '18: ACM Symposium on Neural
  Gaze Detection}{June 03--05, 2018}{Woodstock, NY}
\acmBooktitle{Woodstock '18: ACM Symposium on Neural Gaze Detection,
  June 03--05, 2018, Woodstock, NY}
\acmPrice{15.00}
\acmISBN{978-1-4503-XXXX-X/18/06}

\begin{document}
%\fancyhead{}
%\title{LLM-powered User Interest Exploration for \\Large-scale Recommendation Systems}

%\title{LLMs for User Interest Exploration: A Hybrid Approach}
\title{LLMs for User Interest Exploration in Large-scale Recommendation Systems}

% \renewcommand{\shortauthors}{Kaize, et al.}

% \author{Jianling Wang, Haokai Lu, Yifan Liu, He Ma, Yueqi Wang, Yang Gu, Shuzhou Zhang, Ningren (Peter) Han, Shuchao Bi, Lexi Baugher, Ed Chi and Minmin Chen}
% \affiliation{%
%   \institution{Google DeepMind}
% }
% \email{{jianlingw,haokai,yifanliu,htm,yueqiw,greeness,shuzhouz,peterhan,shuchaobi,baugher,edchi,minminc}@google.com}

\author{Jianling Wang*}\thanks{* indicates Equal Contribution}
\affiliation{%
  \institution{Google DeepMind}
}
\email{jianlingw@google.com}

\author{Haokai Lu*}
\affiliation{%
  \institution{Google DeepMind}
  }
\email{haokai@google.com}

\author{Yifan Liu}
\affiliation{%
  \institution{Google}
  }
\email{yifanliu@google.com}

\author{He Ma}
\affiliation{%
  \institution{Google}
  }
\email{htm@google.com}

\author{Yueqi Wang}
\affiliation{%
  \institution{Google}
  }
\email{yueqiw@google.com}

\author{Yang Gu}
\affiliation{%
  \institution{Google}
  }
\email{greeness@google.com}

\author{Shuzhou Zhang}
\affiliation{%
  \institution{Google}
  }
\email{shuzhouz@google.com}

\author{Ningren Han}
\affiliation{%
  \institution{Google}
  }
\email{peterhan@google.com}

\author{Shuchao Bi}
\affiliation{%
  \institution{Google}
  }
\email{shuchaobi@google.com}

\author{Lexi Baugher}
\affiliation{%
  \institution{Google}
  }
\email{baugher@google.com}

\author{Ed H. Chi}
\affiliation{%
  \institution{Google DeepMind}
  }
\email{edchi@google.com}

\author{Minmin Chen}
\affiliation{%
  \institution{Google DeepMind}
  }
\email{minminc@google.com}

\begin{abstract}
Traditional recommendation systems are subject to a strong feedback loop by learning from and reinforcing past user-item interactions, which in turn limits the discovery of novel user interests. To address this, we introduce a hybrid hierarchical framework combining Large Language Models (LLMs) and classic recommendation models for user interest exploration. The framework controls the interfacing between the LLMs and the classic recommendation models through ``interest clusters'', the granularity of which can be explicitly determined by algorithm designers. It recommends the next novel interests by first representing ``interest clusters'' using language, and employs a fine-tuned LLM to generate novel interest descriptions that are strictly within these predefined clusters. At the low level, it grounds these generated interests to an item-level policy by restricting classic recommendation models, in this case a transformer-based sequence recommender to return items that fall within the novel clusters generated at the high level. We showcase the efficacy of this approach on an industrial-scale commercial platform serving billions of users. Live experiments show a significant increase in both exploration of novel interests and overall user enjoyment of the platform.

\end{abstract}




% <concept_id>10002951.10003317</concept_id>
% <concept_desc>Information systems~Information retrieval</concept_desc>

% \ccsdesc[500]{Information systems~Information retrieval}

\renewcommand{\shortauthors}{Jianling Wang et al.}
\maketitle

\section{Introduction}

Recommendation systems are indispensable in helping users navigate the vast and ever-growing content on the web nowadays. These systems however are often subject to a strong feedback loop \cite{chaney2018algorithmic,mansoury2020feedback}, recommending items similar to a user's past behavior. Classic recommendation systems infer a user's next interest based on their historical interactions. While this can be effective for short-term engagement, it limits users from discovering novel interests, leading to content fatigue. Recent research highlights the importance of \textbf{user interest exploration}~\cite{su2024long,chen2021values,chen2021exploration,mahajan2023pie,song2022show}, aiming to introduce diverse content that goes beyond a user's historical preferences. Effectively introducing novel interests to users are however challenging due to the vast interest space and the high uncertainty of a user's affinity to previously unseen interests \cite{chen2021values,wu2024result}. 

%Prior work \cite{li2023text,li2023gpt4rec,hou2023large,geng2022recommendation} demonstrates the potential of using LLMs for ID-free recommendation by converting user queries and content features into text for generative models. However, the latency of serving LLMs per user request often exceeds the O(100ms) response time expected in production environments, hindering their adoption for large-scale recommendation tasks. Real-world deployment faces significant challenges, including: (1) \textit{Recommendation Object Representation}: LLMs aren't pre-trained or fine-tuned specifically for recommendation tasks. This means they lack deep understanding of specific domains and struggle to keep up with constantly changing data; (2) \textit{Planning Complexity} Managing the vast planning space involved in generating diverse and relevant recommendations; (3) \textit{Alignment and Data Quality}: Ensuring LLM recommendations match both user preferences and platform goals is tricky. 

%Large language models (LLMs), trained on massive web data, possess a comprehensive understanding of the world, demonstrating impressive reasoning and generalization capabilities~\cite{anil2023palm,brown2020language}. 
Recent breakthroughs in Large Language Models (LLMs)~\cite{anil2023palm,brown2020language,touvron2023llama} and other foundation models offer exciting opportunities to revolutionize recommendation systems \cite{bao2023tallrec,dai2023uncovering,lin2023rella,wang2024large,christakopoulou2023large}. The pre-trained world knowledge in these models holds the potential to break recommendation feedback loops by introducing diverse and serendipitous recommendations, addressing the challenge of user interest exploration. While prior work \cite{li2023text,li2023gpt4rec,hou2023large,geng2022recommendation} has demonstrated the potential of using LLMs for recommendation by translating recommendation problems into natural language processing tasks, deploying these approaches in \textit{real-world} industrial recommendation systems remain extremely challenging as: (1) unlike domain-specific recommendation models, LLMs lack deep knowledge of the massive, and rapidly evolving item corpus on industrial-scale online platforms (e.g., more than 500 hours of content are uploaded to YouTube every minute \cite{youtubePress,tubefilter2019}, a new track is uploaded to Spotify every second \cite{mbw}); (2) off-the-shelf LLMs are unaware of the collaborative signals from users, failing to capture domain-specific user behaviors; and (3) the latency and cost of serving LLMs per user request are prohibitively large, cannot meet the O(100ms) response time expected and production Query-Per-Second (QPS) required on industrial recommendation platforms.

\begin{figure}[t]
% \vspace{-0.15in}
\centering
\includegraphics[width=0.35\textwidth]{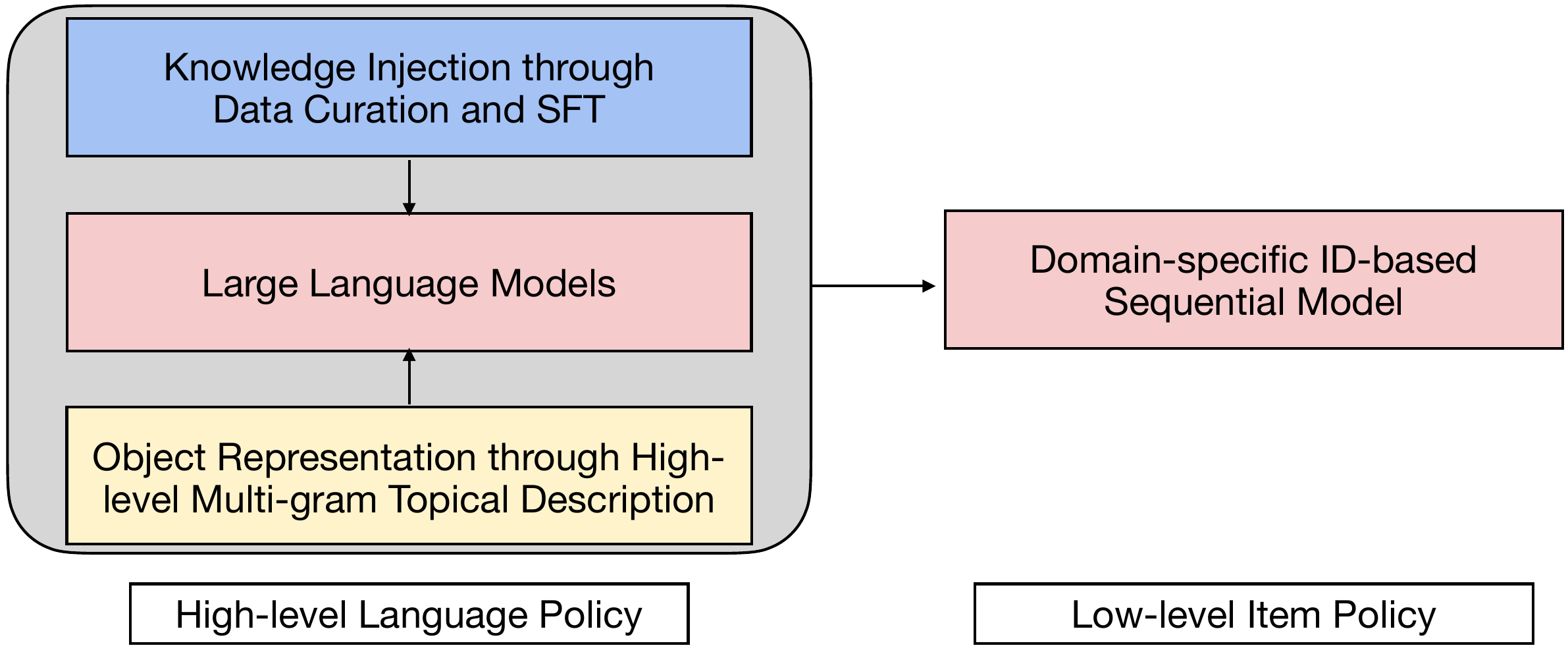}
\vspace{-0.1in}
\caption{LLM-powered hybrid hierarchical planning diagram for user interest exploration.}
\vspace{-0.1in}
\label{fig:illustration}
\end{figure}

To overcome the above challenges, we introduce a \textbf{hybrid hierarchical planning} paradigm combining LLMs and classic recommendation models (as shown in Figure \ref{fig:illustration}) for user interest exploration in large-scale recommendation systems. At the high level, considering the massive number of incoming items in the system, instead of directly predicting the next item, we use LLMs to infer the next novel interest. At the low level, to leverage classic recommendation models with strong personalization, we ground these novel interests to item recommendations by "restricting" conventional transformer-based sequence models \cite{chen2019top,shaw2018self} to items within the "clusters" defined by those novel interests.  
By combining the best of both worlds, the hybrid approach leverages LLMs’ reasoning and generalization capability in exploring user’s novel interests effectively, at the same time bridges the knowledge gap by relying on domain-specific models for actual item recommendation. 
We further perform supervised fine-tuning (SFT) with real-world novel consumption behaviors for in-domain user alignment, and also enable LLMs to perform \textbf{controlled generation}, producing novel interest descriptions that directly match one of the pre-defined clusters. The controlled generation allow algorithm designers to easily define the granularity of the interests generated by LLMs for different applications, which is critical for effectively exploring the interest space.  We find diversification and label balance treatment while curating the SFT data significantly mitigate the long-tailed distribution of LLM generation, and thus improves interest exploration efficiency. To address the LLM inference challenge, we propose to use topically coherent interest clusters with cluster-level descriptions to represent recommendation objects, i.e, both the historical user interests and the recommended next novel interests. By using a small number of historical consumed clusters as high-level user interest, we pre-compute the novel interest transitions offline with LLM bulk inference, which can be then be served online with simple table lookup operations. In summary, we make the following contributions:

\squishlist
%\item We combine the LLM's language understanding with the efficiency of a specialized sequence model for recommendations. This addresses the LLM's lack of domain-specific knowledge;
\item We propose a hybrid hierarchical framework that combines LLM's reasoning and generalization capabilities with classic recommendation models with strong personalization and grounded item corpus knowledge for effective user interest exploration. 
\item We fine-tune LLMs using a diverse and balanced set of novel interest transitions curated from real-world user interactions for controlled generation and user behavior alignment, to ensure LLMs generate novel interests that match one of the predefined interest "clusters" and align with actual user behaviors.
\item We propose to adopt topical clusters instead of items to represent user's high-level interests. The coarser representation allows us to limit the length of historical cluster sequence used to represent dynamic user interests and move the expensive LLM inference to offline stage, making it feasible to serve LLM generated novel interest transitions online.
\item We validate our method through live experiments on a large commercial recommendation platform with billions of users. The results clearly show our approach successfully expands user interests while boosting user enjoyment of the platform demonstrated through more active users with longer dwell time.
\squishend

\section{Related Work}
\smallskip
\textbf{LLMs for Recommendation Systems}. 
Application of LLMs to recommendation systems is a rapidly growing research area. Some studies explore using LLMs directly for generating recommendations \cite{bao2023tallrec,dai2023uncovering,geng2023vip5,hou2023large,li2023gpt4rec,liu2023chatgpt}, while others focus on augmenting traditional recommendation models with LLM-powered feature engineering \cite{hou2023learning,yu2021tiny} or enriched user/item representations \cite{li2023taggpt,liu2023first,xi2023towards}. The computational cost of LLMs however presents a critical challenge. Directly using them for large-scale retrieval is expensive and hinders adoption. Wang et al. \cite{wang2024large} use LLMs as data augmenters for conventional recommendation systems during training, to improve model performance without additional serving cost. 
% Similarly, this approach aims to make LLMs viable for industrial-scale recommendation systems with strict query-per-second (QPS) requirements. 
Different from prior work, we focus on directly incorporating LLM-generated content to break the feedback loop, aiming for more diverse and serendipitous recommendations while maintaining efficiency.

\smallskip
\noindent\textbf{User Interest Exploration}. 
Prior research has established the benefits of User Interest Exploration in recommendation systems, demonstrating its ability to expand user preferences and enhance long-term engagement \cite{chen2021values,chen2021exploration, su2024long}. However, a key challenge lies in the inherent closed-loop nature of existing systems~\cite{chaney2018algorithmic,mansoury2020feedback,chen2021values}. Training data is primarily derived from past user-item interactions, limiting the system's ability to explore truly novel interests. While methods like PIE \cite{mahajan2023pie} address this to certain extent through user-creator affinity and online bandit formulations, they remain confined by the system's internal knowledge \cite{chen2021values}. Our work introduces a novel approach that integrates world knowledge from LLMs to overcome these limitations. % and meanwhile facilitate more comprehensive user interest exploration.

\section{Method}
In this section, we introduce the hybrid hierarchical planning paradigm and the LLM fine-tuning process designed to enable controlled generation and user behavior alignment, to apply LLMs in real-world large-scale recommenders.

\subsection{Preliminaries}
\label{sec:preliminary}
% Let $\mathcal{U}$ and $\mathcal{I}$ the represent the user and item set. 
The sheer number of items and the constant influx of new items on online platforms make LLM planning at the individual item-level infeasible. Instead, we leverage the planning capabilities of LLMs at the item interest-level to reduce the planning space. A prerequisite for efficient hierarchical planning is a set of high-quality \textit{item interest clusters}, where items within each cluster are topically coherent. Following the same procedure as in \cite{chang2024cluster}, we group items into $N$ traffic-weighted equal sized \textbf{clusters} based on their topical coherence, a method proven to scale well to the magnitude of our problem. To create these clusters, we first represent each item as a 256-dimensional embedding based on its metadata (title, hashtags, etc.) and content (frames and audio). Then, we connect items in a graph based on their similarity and cluster it into traffic-balanced clusters. This clustering process is repeated multiple times to create a 4-level tree structure, with each item associated with different tree levels. Higher-level clusters represent broader topics, while lower-level clusters represent more specific ones. These clusters in each level, denoted by $\mathcal{C}^l=\{c_1^l, c_2^l, ..., c_{M_l}^l\}, l=1, 2, 3, 4$, represent different user interests, with each cluster linked to a \textbf{set of keywords} describing its theme. Here $M_l$ denotes the number of clusters within level $l$, with $M_4 > M_3 > M_2 > M_1$. Each item belongs to a single interest cluster in each level. As discussed in \cite{chang2024cluster}, we focus on level-2 clusters to balance granularity and feasible planning space\footnote{To reduce notation complexity, we drop level $l$ in cluster moving forward.}.

%Given a sequence of previously watched videos, plan the next language-based novel interests through LLMs.

% \haokai{TODO: consider remove this if there is no space.}
% Large recommendation systems in production are often built in multiple stages \cite{covington2016deep,ma2020off}: in which the first stage involves various retrieval models that identify candidate items from the overall corpus, the second stage scores and ranks these item, while the last stage packs them with consideration for diversity and specific business objectives. We focus on the \textbf{retrieval stage} because it's crucial for introducing items that align with new user interests. 

\subsection{Hybrid Hierarchical Planning}
%In this section, we propose a hierarchical planning framework to explore user's interests by recommending novel items. 
The hybrid approach combines a LLM to produce a language policy that generates novel interests on the high-level, and a classic recommendation models to produce an item policy that grounds these language-based interests to the low-level item space. 
% On the high level, the framework adopts a language policy to generate language-based novel interests first; and on the low level, it adopts an item-based policy to convert these language-based interests to items that are within the interest clusters. 
Such a hybrid approach combines the strength of LLMs in reasoning and generalization, and domain-specific recommendation models in handling item dynamics and enhanced personalization.       

\begin{figure}[t]
\vspace{-0.15in}
\centering
\includegraphics[width=0.45\textwidth]{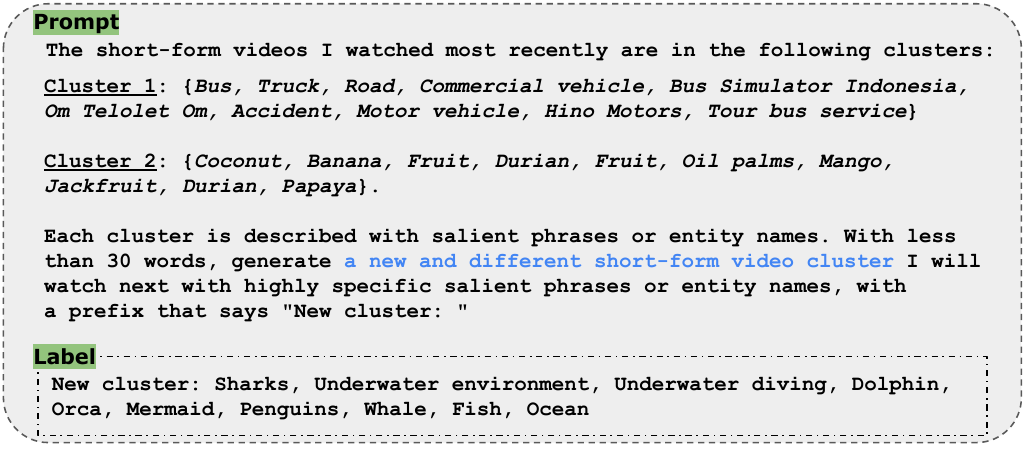}
\vspace{-0.1in}
\caption{Prompt for Novel Interest Prediction when $K=2$.}
\vspace{-0.1in}
\label{fig:prompt}
\end{figure}

\smallskip
\noindent\textbf{High-level language policy.} 
Given the historical user interest representation by language, we first use LLM to learn a high-level language policy that generates novel user interests. Instead of using item descriptions to represent users, we propose to adopt cluster descriptions (i.e., a set of keywords) to represent user's consumption history, i.e., a user's historical interest is represented as a sequence of her $K$ most recent interacted unique clusters, with each cluster represented by its description. Specifically, with a user's previously consumed unique clusters, we can ask LLM to generate the next novel interest with the prompt illustrated in Figure \ref{fig:prompt}.    

%Instead of using LLM to directly perform the next recommendation in the item space, we abstract a level up and given historical viewer interest representation by language, we plan the next novel interest through LLM. 
%Specifically, a user's current interests are represented as a sequence of their N most recent interacted clusters. The top 10 topical words from each cluster in the user's interaction history provide high-quality representations of their interests at a broad level. Using these topical words from different clusters to construct prompts, we ask the LLM to generate the user's next novel interests, as illustrated in Figure \ref{fig:prompt}.
% At serving time, we dynamically and personally map users to an interest bucket with thompson sampling, look up the precomputed interest transition mapping, and serve the lookup results

\smallskip
\noindent\textbf{Practical implication.} One major challenge of deploying LLM to industrial-scale recommendation system lies in its prohibitively high inference cost failing to meet latency and QPS requirements. Empirically, we find that relying on a small number of historical clusters to represent each user (e.g., $K=2$) can effectively balance representation granularity and computation efficiency. In our experiment, the level-2 clustering produced 761 clusters.  We can therefore enumerate all $761*761 = 579,121$ cluster pairs and perform a batch inference with LLM to obtain novel interests for each cluster pair under just a few hours. These novel interests, together with the input cluster pairs, could be stored in a table.  During \textit{online serving} as a new user request comes in, we first represent the user by sampling $K=2$ items from their watch history\footnote{30-day user history are used in sampling, and items with high-quality interactions are more likely to be sampled.}, and convert them into the cluster pair for lookup to determine the recommended novel interest cluster. %,  with simple lookup operations to meet the latency requirement.   
%a To address that, we select a small number of historical clusters consumed by the users (e.g., $N=2$) to 

\smallskip
\noindent\textbf{Low-level item policy.} Once the language based novel user interest is obtained, the next step is to convert it to item-level recommendation policy. A straight-forward approach is to rely on search engine \cite{li2023gpt4rec} to retrieve the most relevant items according to the keywords of the novel interest. The search results however often lack personalization since these language based novel interests can still be broad and lack specificity. To enhance personalization, we propose to reuse domain specific recommendation models, specifically \textit{transformer-based sequential recommender} model \cite{chen2019top,shaw2018self}, but restrict the items to clusters prescribed by the language based novel interests. Specifically, we follow the two steps: (i) map the generated novel interests to cluster ID space, and (ii) restrict the original item level softmax policy on these cluster IDs, to retrieve items \textbf{only} from these clusters.  

\begin{figure}[t]
\vspace{-0.1in}
\centering
\includegraphics[width=0.28\textwidth]{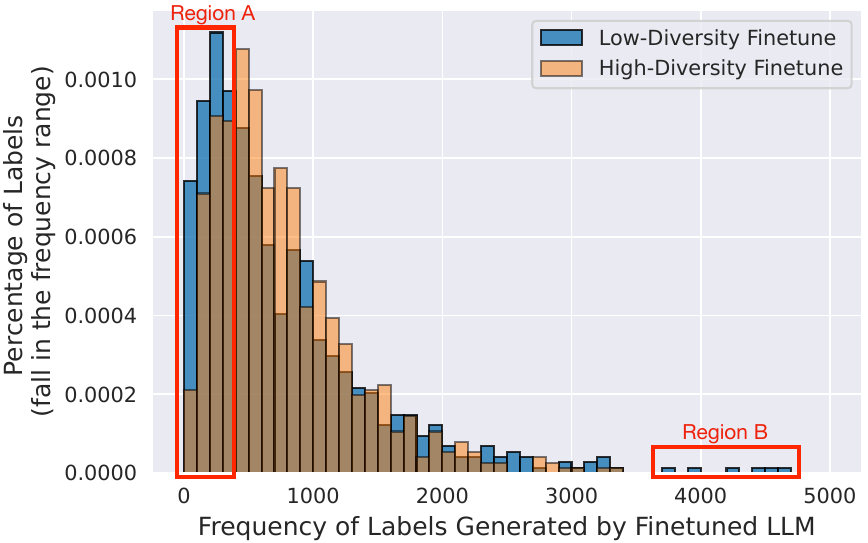}
\vspace{-0.1in}
\caption{Label (i.e., generated by fine-tuned LLM) Distribution: X-axis represents label frequency; Y-axis represents the percentage of labels within each frequency range.}
\label{fig:distribution}
\vspace{-0.1in}
\end{figure}

\smallskip
\noindent\textbf{Controlled generation.} The remaining challenges are how to specify the granularity of the LLM generated interests and map the generated novel interests to cluster IDs. Free-form responses from LLMs can be arbitrary, which are unlikely to be directly matched to the predefined cluster descriptions. We control the granularity of the generation through the hierarchical clustering and picking the cluster level explained in Section ~\ref{sec:preliminary}.  Furthermore, appropriate finetuning as detailed in Section \ref{sec:finetune} enables LLMs to speak the language of interest clusters, producing cluster descriptions that exactly match one of the predefined clusters.

%We apply the RNN-based sequence model \cite{chen2019top}
%Task: given language-based novel interests, plan the next novel Shorts video to viewers (grounding language output to video space).
%Approach: we propose to map the generated interests back to cluster ID space so that these cluster IDs can be used as restrict tokens to restrict sequence to videos within these clusters. 

%One major benefit of adopting sequence restrict for video level nomination is that the main discovery sequence is highly optimized for personalization given that the cluster is still at L2 level. 

%One challenge of mapping the Gemini’s response to cluster ID is that the response can be arbitrary and does not directly map to the descriptions of clusters. We address the issue of mapping language output from Gemini to cluster ID with finetuning.

\begin{figure*}[!t]
\vspace{-0.2in}
    \graphicspath{{figures/}}
    %\centering
    %\hspace{-1cm}
    \scalebox{0.88}{
    \subfigure[Control Generation Capability \& Alignment Learning]
    {
    \includegraphics[height=0.215\textwidth]{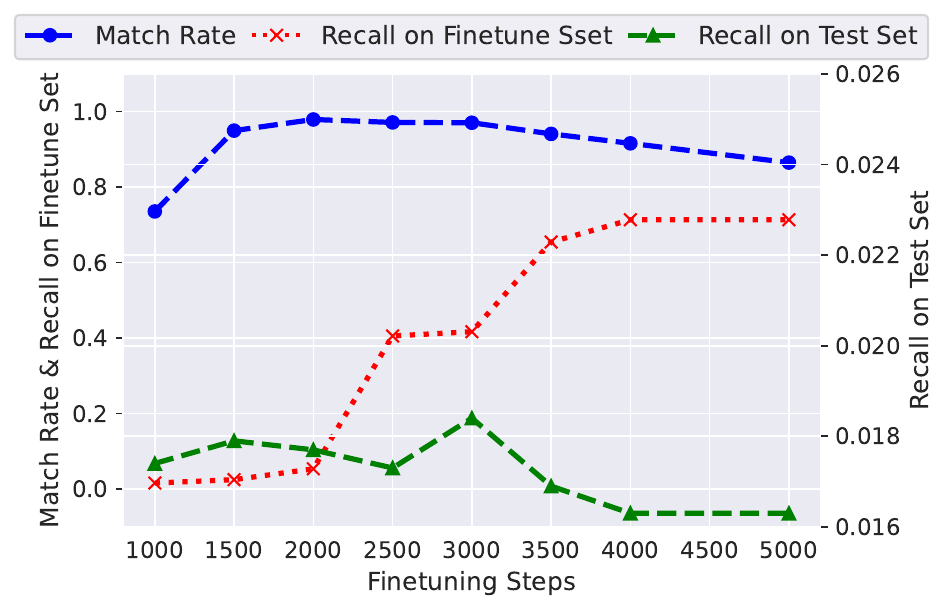}
    }
    \hspace{0.2cm}
    \subfigure[Novelty (x-axis) and Quality (y-axis) Comparison]
    {
    \includegraphics[height=0.215\textwidth]{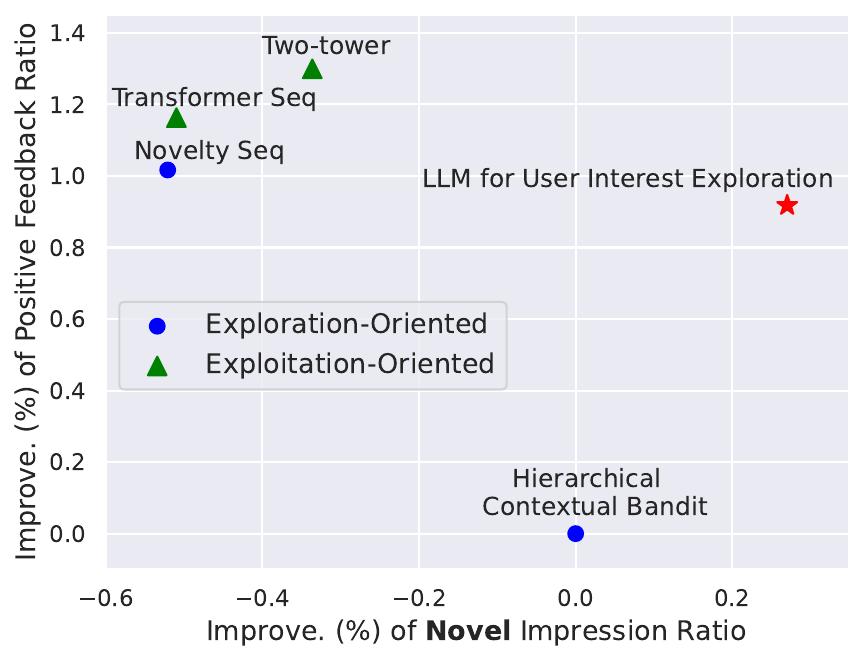}
    }
    \hspace{0.2cm}
    \subfigure[Improvement of UCI@N]
    {
    \includegraphics[height=0.215\textwidth]{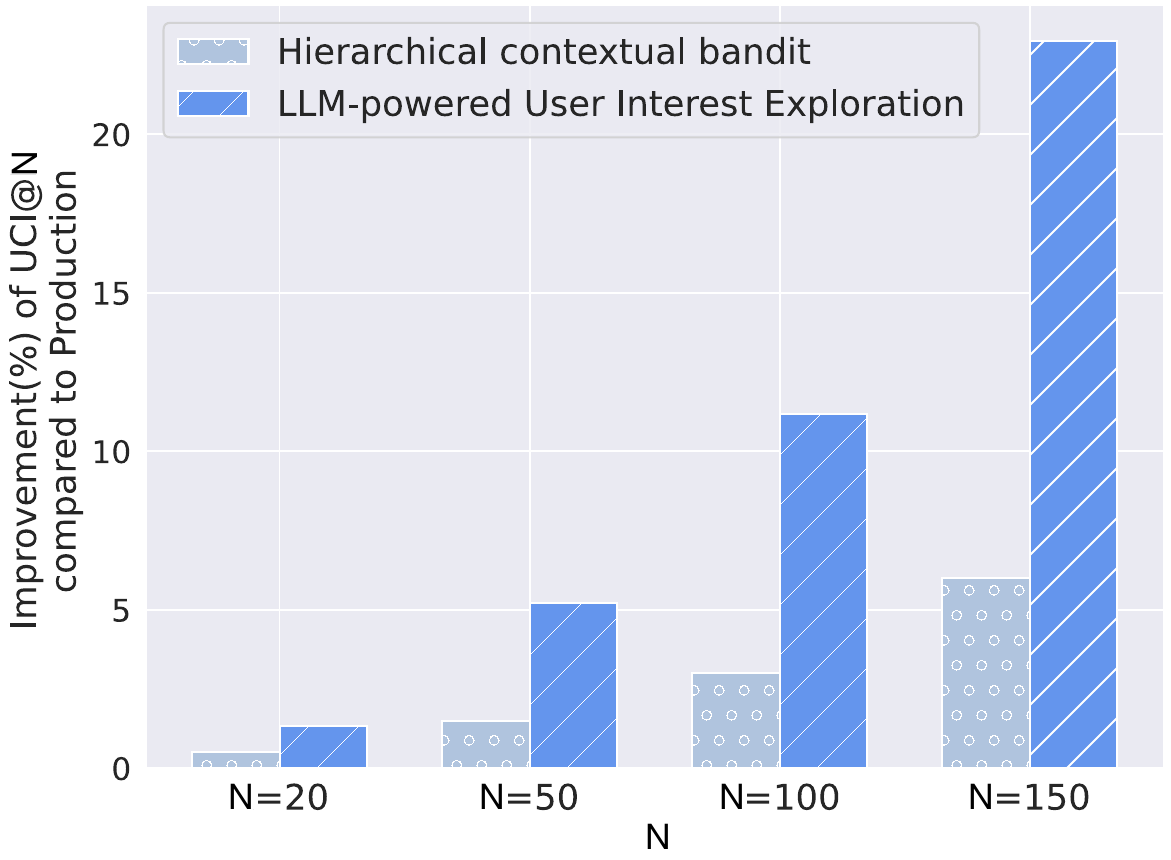}
    }
    }
    \vspace{-0.15in}
    \caption{(a) Model Finetuning Process. (b) and (c) Comparison between different recommenders in live experiments.}%
    \vspace{-0.1in}
    \label{fig:finetune_process}
    \label{fig:supi}
\end{figure*}

\subsection{Fine-Tuning for User Behavior Alignment}
\label{sec:finetune}
LLMs trained on massive publicly available data on the Internet contains rich global knowledge, it however lacks capabilities to perform: 1) controlled generation (i.e., generate in the interest cluster space) and 2) domain-specific user behavior alignment. We propose to inject these domain specific knowledge through supervised fine-tuning using a dataset curated from real user watch histories on the commercial platform. The quality of data used for fine-tuning is thus crucial for its success. 
%To achieve these goals, we fine-tune LLM using a dataset curated from real user watch histories on our commercial platform. This dataset is sufficiently diverse to encompass all interest clusters represented in the labels.
%The quality of data used for fine-tuning is crucial for this project’s success. %By (1) incorporating real user behavior with transitions between different clusters; and (2) ensuring the data is diverse, covering all clusters evenly, we can mitigate the long-tailed distribution issue in the model's generated clusters. This, in turn, not only mitigates the feedback loop effect in behavior data but also improves overall offline accuracy, reflected in higher recall (i.e., hit rate) (see Section \ref{sec:offline}). 

\smallskip
\noindent\textbf{Diversified Data Curation.} Take K=2 as an example. Each fine-tuning data sample, denoted as $[(C_1, C_2), C_L]$, consists of a cluster pair $(C_1, C_2)$ to form the prompt and the subsequent novel cluster $C_L$ as the finetune label. By definition of novelty, $C_L$ must be different from $C_1$ and $C_2$. Initially, we gathered approximately 250K $[(C_1, C_2), C_L]$ data samples from our log, focusing solely on high quality interactions\footnote{In other words, these samples are novel interest transitions existed in the log where a user was successfully introduced a novel interest.}. These samples are then grouped by their labels, and the top-10 most frequently occurring cluster pairs are selected for each label, forming the final data samples which is diverse and cover all the labels. Following these steps, we obtain $761*10 =7,610$ data samples (10 per label cluster) and proceed to perform supervised fine-tuning for the LLM using these samples.

In Figure \ref{fig:distribution}, we plot the distributions of interest clusters generated by fine-tuned LLMs on the 579,121 context cluster pairs. 
%The X-axis represents the number of times LLM predicts a label, while Y-axis represents the percentage of labels falling in the frequency range. 
When using finetuning data of low diversity, where we randomly select 7,610 transitions and their corresponding subsequent clusters from the initial $250K$ to form the data, the fine-tuned LLM generates interests that are highly-skewed, where a few generated clusters have very high frequencies (depicted as Region B). When we increase the diversity of our fine-tuning data, these dominant labels disappear, and the number of generated clusters with very low frequency (depicted in Region A) also decreases. %This means we have fewer labels that receive extremely low frequencies.  
Ensuring the fine-tuning data covers all clusters evenly allows us to address the long-tailed distribution issue in the model's generated clusters.  This treatment not only mitigates the feedback loop effect in behavior data but also enhances overall user satisfaction as shown in Section \ref{sec:live_exp}.

\smallskip
\noindent\textbf{Control Generation Capability \& User Behavior Alignment.}
The number of fine-tuning steps determines the balance between the LLM’s global  and task-specific knowledge. Our fine-tuning process has two main objectives: (1) \textbf{controlling} LLM generation to speak the language of interest clusters. We evaluate the \textit{match rate} of the generation from the fine-tuned LLM to determine if the output matches exactly with one of the cluster descriptions; and (2) \textbf{aligning} with real-world user transitions, measured by comparing the fine-tuned LLM's output with the successful user interest transition in both fine-tuning and test set to compute \textit{recall}. A higher recall indicates LLM learning the domain specific novel transitions from the fine-tuned data,  and aligning with user behaviors. % from the real-world fine-tuning dataset.

In Figure \ref{fig:finetune_process} (a), with batch size of 16, we illustrate the changes in match rate and recall as the fine-tuning steps progress. We note that formatting learning, i.e., learning the language of intrest clusters, kicks in first, peaking at around 2,000 steps. With a high match rate (over 99\%), we can efficiently map the generation to cluster ID space and restrict the original item-level softmax policy on these clusters. Subsequently, the model begins to align with user behaviors, resulting in a significant increase in recall (on the fine-tuned set). Moreover, we find that the recall for a separate test set increases following transition alignment, reaching its peak at around 3,000 steps before gradually decreasing. Therefore, we select models fine-tuned with 3,000 steps. Note that the recall on the test set is much lower than that of the fine-tuning set, indicating LLM is still relying heavily on its global knowledge instead of memorizing interest transitions in the log while generating novel interests. %We also ensure that the model generates new clusters distinct from historical ones, with only 1\% of cases repeating clusters in the prompt. .

\section{Live Experiments}
\label{sec:live_exp}
%In this section, we examine how effectively a hierarchical planning with LLM design improves user interest exploration. 
\subsection{Experimental Setup}
We conduct a series of live experiments on a \textit{commercial short-form video recommendation platform that serves billions of users}. Our experiments are conducted with \textbf{Gemini-Pro} \cite{team2023gemini}, but the same fine-tuning process and pipeline can be readily adapted to other LLMs. %In our initial implementation, we use a small number of clusters (K=2) from a user's watch history to balance the level of detail with computational efficiency. 
We set the number of historical clusters for LLM inference $K=2$, it however can easily scale to accommodate larger numbers in future iterations with a sparse table.

\smallskip
\noindent\textbf{Baseline.} 
We compare the proposed method to existing production models: (1) \textbf{Exploration-oriented} models include: a \textit{Novelty-enhanced sequence recommender} ~\cite{chen2021values} trained with labels from both positive and novel items whose clusters have not appeared in user's consumption history before; \textit{Hierarchical contextual bandit} \cite{song2022show} based on the hierarchical clusters introduced in \ref{sec:preliminary} to explore user's interests through a tree-based LinUCB to obtain the next clusters, from which the sequential model is then used to restrict the retrieval to items.
%; \textit{Neural linear bandit}-based DNN model \cite{su2024long} to predict the next novel cluster, from which the same sequential model is then used for item-level policy, similarly as hierarchical contextual bandit. 
Although these models are tailored to explore user interests, they are trained on interest transitions existing in the system and therefore are still subject to the feedback loop. (2) \textbf{Exploitation-oriented} models include a regular \textit{two-tower} model \cite{yang2020mixed} and \textit{transformer-based} \cite{chen2019top,shaw2018self} sequential model trained on all positive user feedback. Our live experimental results demonstrate our proposed method can lead to recommendation which are more novel and in better quality compared to these existing models. 

%We compare with the following baselines which are already deployed in production: Type(1) \textbf{Novelty-enhanced sequence recommender} ~\cite{chen2021values}. Such model is trained with labels from both positive and novel items whose clusters have not appeared in user's consumption history before. (2) \textbf{Hierarchical contextual bandit} \cite{song2022show} based on the hierarchical clusters introduced in \ref{sec:preliminary}. Such model explores user's interests through a tree-based LinUCB to obtain the next clusters, from which the sequential model is then used to restrict the retrieval to items. (3) \textbf{Neural linear bandit}-based DNN model \cite{su2024long} to predict the next novel cluster, from which the same sequential model is then used for item-level policy, similarly as hierarchical contextual bandit.(4) Two-tower model and transformer-based sequential model trained for exploitation. Though this models are tailored for extend user interest, they are trained on interest transitions existing in the system and therefore still fall into the feedback loop. (5)  And our live experimental results demonstrate our proposed method can lead to recommendation which are more novel and in better quality compared to the existing models. 

\subsection{Results and Analysis}

\noindent\textbf{Novelty and Quality.} In Figure \ref{fig:supi} (b), we compare the proposed method with various baseline models currently in production. Using the performance of Hierarchical contextual bandit \cite{song2022show} as the reference, we measure the improvement of the other models. Specifically, we plot the increase in ratio of novel impressions (considering only impressions from interest clusters the user has never interacted with) to highlight recommendation novelty (x-axis), and the increase in positive feedback rate to demonstrate recommendation quality (y-axis). The proposed method recommends more novel items compared to all the baseline methods (to the right on x-axis). Additionally, it achieves much better quality than existing exploration-oriented methods, comparable to exploitation-oriented methods (high on x-axis). In other words, the proposed method presents an effective approach to introduce users to novel interests that are of interests to the user. % user interest exploration while preserving effective exploitation. 

\smallskip
\noindent\textbf{User Interest Exploration.} To measure if the recommenders encourage users to explore new interests, we use a metric called \textbf{UCI@N}, which tracks the number of users who have consumed items from N \underline{\textbf{u}}nique \underline{\textbf{c}}lustered \underline{\textbf{i}}nterests within the past 7 days. Higher UCI@N indicates more users are consuming N interests. By monitoring UCI@N for different values of N (20 to 200), we can gauge the effectiveness of our system for user interest exploration. Figure \ref{fig:supi}(c) summarizes the improvement of our method compared to Hierarchical Contextual Bandit, to evaluate its effectiveness in user interest exploration. Notably, our proposed method shows very significant improvement compared to the prominent exploratory model currently deployed in production for different values of $N$.

\begin{figure}[h]
\vspace{-0.15in}
    %\graphicspath{{figures/}}
    \centering
    \subfigure[Overall Watch Time]
    {
    \includegraphics[width=0.43\columnwidth]{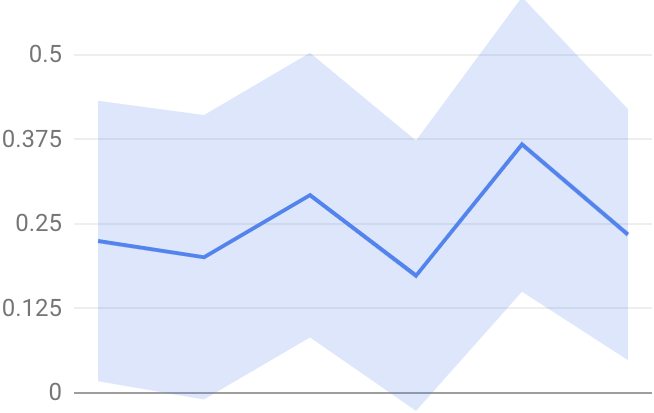}
    }
    \hspace{0.05in}
    \subfigure[Number of Users watch >= 10min]
    {
    \includegraphics[width=0.41\columnwidth]{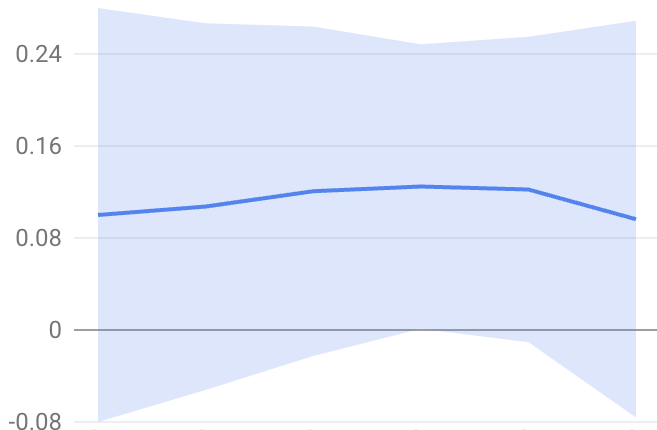}
    }
    \vspace{-0.15in}
    \caption{The proposed method drives user growth.}
    \label{fig:live}
    \vspace{-0.15in}
\end{figure}

\smallskip
\noindent\textbf{User Growth.} At the same time, we monitor increase in overall watch time and number of active users who had total watch time $>=$ 10 minute (in Figure \ref{fig:live}), to measure user growth on the short-form video platform. %  illustrates the user metrics from our live experiments. 
The x-axis represents the experiment periods (the exact dates are redacted), and the y-axis shows the relative percentage difference between the experiment and control, which excludes the proposed system. Our method successfully broadens user interest by recommending diverse and novel content, with user growth. This underscores the quality and relevance of the recommended novel content.

\section{Conclusion}
We present a hybrid approach to leverage LLMs for user interest exploration. % framework that leverages LLMs for large-scale recommendation systems.
It combines the strength of LLMs in reasoning and generalization, and the grounding of classic recommendation models.
We showcase a successful recipe to inject domain-specific recommendation knowledge to LLMs for controlled generation and user behavior alignment. 
% Our approach first generates language-based novel interests, and then translates them into item-level recommendations by restricting candidate selection to items within clusters defined by these novel interests. 
Extensive testing on a commercial platform with billions of users has yielded significant improvements in both exploration of novel interests and user growth.
The future work will focus on 
% incorporating domain-specific knowledge to enhance user understanding at an abstract level, and
taking long-term effects into account to further improve hierarchical planning with LLMs for recommendation systems.

\balance
\bibliographystyle{ACM-Reference-Format}
\bibliography{sample_bib}


\begin{thebibliography}{35}

%%% the Web address.

\ifx \showCODEN    \undefined \def \showCODEN     #1{\unskip}     \fi
\ifx \showDOI      \undefined \def \showDOI       #1{#1}\fi
\ifx \showISBNx    \undefined \def \showISBNx     #1{\unskip}     \fi
\ifx \showISBNxiii \undefined \def \showISBNxiii  #1{\unskip}     \fi
\ifx \showISSN     \undefined \def \showISSN      #1{\unskip}     \fi
\ifx \showLCCN     \undefined \def \showLCCN      #1{\unskip}     \fi
\ifx \shownote     \undefined \def \shownote      #1{#1}          \fi
\ifx \showarticletitle \undefined \def \showarticletitle #1{#1}   \fi
\ifx \showURL      \undefined \def \showURL       {\relax}        \fi
% invisible to TeX
\providecommand\bibfield[2]{#2}
\providecommand\bibinfo[2]{#2}
\providecommand\natexlab[1]{#1}
\providecommand\showeprint[2][]{arXiv:#2}

\bibitem[\protect\citeauthoryear{Anil, Dai, Firat, Johnson, Lepikhin, Passos, Shakeri, Taropa, Bailey, Chen, et~al\mbox{.}}{Anil et~al\mbox{.}}{2023}]%
        {anil2023palm}
\bibfield{author}{\bibinfo{person}{Rohan Anil}, \bibinfo{person}{Andrew~M Dai}, \bibinfo{person}{Orhan Firat}, \bibinfo{person}{Melvin Johnson}, \bibinfo{person}{Dmitry Lepikhin}, \bibinfo{person}{Alexandre Passos}, \bibinfo{person}{Siamak Shakeri}, \bibinfo{person}{Emanuel Taropa}, \bibinfo{person}{Paige Bailey}, \bibinfo{person}{Zhifeng Chen}, {et~al\mbox{.}}} \bibinfo{year}{2023}\natexlab{}.
\newblock \showarticletitle{Palm 2 technical report}.
\newblock \bibinfo{journal}{\emph{arXiv preprint arXiv:2305.10403}}.
\newblock


\bibitem[\protect\citeauthoryear{Bao, Zhang, Zhang, Wang, Feng, and He}{Bao et~al\mbox{.}}{2023}]%
        {bao2023tallrec}
\bibfield{author}{\bibinfo{person}{Keqin Bao}, \bibinfo{person}{Jizhi Zhang}, \bibinfo{person}{Yang Zhang}, \bibinfo{person}{Wenjie Wang}, \bibinfo{person}{Fuli Feng}, {and} \bibinfo{person}{Xiangnan He}.} \bibinfo{year}{2023}\natexlab{}.
\newblock \showarticletitle{Tallrec: An effective and efficient tuning framework to align large language model with recommendation}.
\newblock \bibinfo{journal}{\emph{arXiv preprint arXiv:2305.00447}}.
\newblock


\bibitem[\protect\citeauthoryear{Blog}{Blog}{2023}]%
        {youtubePress}
\bibfield{author}{\bibinfo{person}{Youtube~Official Blog}.} \bibinfo{year}{2023}\natexlab{}.
\newblock \bibinfo{booktitle}{\emph{YouTube by the Number}}.
\newblock
\urldef\tempurl%
\url{https://blog.youtube/press/}
\showURL{%
Retrieved January, 2023 from \tempurl}


\bibitem[\protect\citeauthoryear{Brown, Mann, Ryder, Subbiah, Kaplan, Dhariwal, Neelakantan, Shyam, Sastry, Askell, et~al\mbox{.}}{Brown et~al\mbox{.}}{2020}]%
        {brown2020language}
\bibfield{author}{\bibinfo{person}{Tom Brown}, \bibinfo{person}{Benjamin Mann}, \bibinfo{person}{Nick Ryder}, \bibinfo{person}{Melanie Subbiah}, \bibinfo{person}{Jared~D Kaplan}, \bibinfo{person}{Prafulla Dhariwal}, \bibinfo{person}{Arvind Neelakantan}, \bibinfo{person}{Pranav Shyam}, \bibinfo{person}{Girish Sastry}, \bibinfo{person}{Amanda Askell}, {et~al\mbox{.}}} \bibinfo{year}{2020}\natexlab{}.
\newblock \showarticletitle{Language models are few-shot learners}. In \bibinfo{booktitle}{\emph{NeurIPS}}.
\newblock


\bibitem[\protect\citeauthoryear{Chaney, Stewart, and Engelhardt}{Chaney et~al\mbox{.}}{2018}]%
        {chaney2018algorithmic}
\bibfield{author}{\bibinfo{person}{Allison~JB Chaney}, \bibinfo{person}{Brandon~M Stewart}, {and} \bibinfo{person}{Barbara~E Engelhardt}.} \bibinfo{year}{2018}\natexlab{}.
\newblock \showarticletitle{How algorithmic confounding in recommendation systems increases homogeneity and decreases utility}. In \bibinfo{booktitle}{\emph{RecSys}}.
\newblock


\bibitem[\protect\citeauthoryear{Chang, Meng, Ma, Chang, Gu, Peng, Feng, Zhang, Bi, Chi, et~al\mbox{.}}{Chang et~al\mbox{.}}{2024}]%
        {chang2024cluster}
\bibfield{author}{\bibinfo{person}{Bo Chang}, \bibinfo{person}{Changping Meng}, \bibinfo{person}{He Ma}, \bibinfo{person}{Shuo Chang}, \bibinfo{person}{Yang Gu}, \bibinfo{person}{Yajun Peng}, \bibinfo{person}{Jingchen Feng}, \bibinfo{person}{Yaping Zhang}, \bibinfo{person}{Shuchao Bi}, \bibinfo{person}{Ed~H Chi}, {et~al\mbox{.}}} \bibinfo{year}{2024}\natexlab{}.
\newblock \showarticletitle{Cluster Anchor Regularization to Alleviate Popularity Bias in Recommender Systems}. In \bibinfo{booktitle}{\emph{Companion Proceedings of the ACM on Web Conference 2024}}. \bibinfo{pages}{151--160}.
\newblock


\bibitem[\protect\citeauthoryear{Chen}{Chen}{2021}]%
        {chen2021exploration}
\bibfield{author}{\bibinfo{person}{Minmin Chen}.} \bibinfo{year}{2021}\natexlab{}.
\newblock \showarticletitle{Exploration in recommender systems}. In \bibinfo{booktitle}{\emph{RecSys}}.
\newblock


\bibitem[\protect\citeauthoryear{Chen, Beutel, Covington, Jain, Belletti, and Chi}{Chen et~al\mbox{.}}{2019}]%
        {chen2019top}
\bibfield{author}{\bibinfo{person}{Minmin Chen}, \bibinfo{person}{Alex Beutel}, \bibinfo{person}{Paul Covington}, \bibinfo{person}{Sagar Jain}, \bibinfo{person}{Francois Belletti}, {and} \bibinfo{person}{Ed~H Chi}.} \bibinfo{year}{2019}\natexlab{}.
\newblock \showarticletitle{Top-k off-policy correction for a REINFORCE recommender system}. In \bibinfo{booktitle}{\emph{WSDM}}.
\newblock


\bibitem[\protect\citeauthoryear{Chen, Wang, Xu, Le, Sharma, Richardson, Wu, and Chi}{Chen et~al\mbox{.}}{2021}]%
        {chen2021values}
\bibfield{author}{\bibinfo{person}{Minmin Chen}, \bibinfo{person}{Yuyan Wang}, \bibinfo{person}{Can Xu}, \bibinfo{person}{Ya Le}, \bibinfo{person}{Mohit Sharma}, \bibinfo{person}{Lee Richardson}, \bibinfo{person}{Su-Lin Wu}, {and} \bibinfo{person}{Ed Chi}.} \bibinfo{year}{2021}\natexlab{}.
\newblock \showarticletitle{Values of user exploration in recommender systems}. In \bibinfo{booktitle}{\emph{RecSys}}.
\newblock


\bibitem[\protect\citeauthoryear{Christakopoulou, Lalama, Adams, Qu, Amir, Chucri, Vollucci, Soldo, Bseiso, Scodel, et~al\mbox{.}}{Christakopoulou et~al\mbox{.}}{2023}]%
        {christakopoulou2023large}
\bibfield{author}{\bibinfo{person}{Konstantina Christakopoulou}, \bibinfo{person}{Alberto Lalama}, \bibinfo{person}{Cj Adams}, \bibinfo{person}{Iris Qu}, \bibinfo{person}{Yifat Amir}, \bibinfo{person}{Samer Chucri}, \bibinfo{person}{Pierce Vollucci}, \bibinfo{person}{Fabio Soldo}, \bibinfo{person}{Dina Bseiso}, \bibinfo{person}{Sarah Scodel}, {et~al\mbox{.}}} \bibinfo{year}{2023}\natexlab{}.
\newblock \showarticletitle{Large language models for user interest journeys}.
\newblock \bibinfo{journal}{\emph{arXiv preprint arXiv:2305.15498}} (\bibinfo{year}{2023}).
\newblock


\bibitem[\protect\citeauthoryear{Dai, Shao, Zhao, Yu, Si, Xu, Sun, Zhang, and Xu}{Dai et~al\mbox{.}}{2023}]%
        {dai2023uncovering}
\bibfield{author}{\bibinfo{person}{Sunhao Dai}, \bibinfo{person}{Ninglu Shao}, \bibinfo{person}{Haiyuan Zhao}, \bibinfo{person}{Weijie Yu}, \bibinfo{person}{Zihua Si}, \bibinfo{person}{Chen Xu}, \bibinfo{person}{Zhongxiang Sun}, \bibinfo{person}{Xiao Zhang}, {and} \bibinfo{person}{Jun Xu}.} \bibinfo{year}{2023}\natexlab{}.
\newblock \showarticletitle{Uncovering ChatGPT's Capabilities in Recommender Systems}.
\newblock \bibinfo{journal}{\emph{arXiv preprint arXiv:2305.02182}}.
\newblock


\bibitem[\protect\citeauthoryear{{Gemini Team Google}}{{Gemini Team Google}}{2023}]%
        {team2023gemini}
\bibfield{author}{\bibinfo{person}{{Gemini Team Google}}.} \bibinfo{year}{2023}\natexlab{}.
\newblock \showarticletitle{Gemini: A family of highly capable multimodal models}.
\newblock \bibinfo{journal}{\emph{arXiv preprint arXiv:2312.11805}} (\bibinfo{year}{2023}).
\newblock


\bibitem[\protect\citeauthoryear{Geng, Liu, Fu, Ge, and Zhang}{Geng et~al\mbox{.}}{2022}]%
        {geng2022recommendation}
\bibfield{author}{\bibinfo{person}{Shijie Geng}, \bibinfo{person}{Shuchang Liu}, \bibinfo{person}{Zuohui Fu}, \bibinfo{person}{Yingqiang Ge}, {and} \bibinfo{person}{Yongfeng Zhang}.} \bibinfo{year}{2022}\natexlab{}.
\newblock \showarticletitle{Recommendation as language processing (rlp): A unified pretrain, personalized prompt \& predict paradigm (p5)}. In \bibinfo{booktitle}{\emph{ResSys}}.
\newblock


\bibitem[\protect\citeauthoryear{Geng, Tan, Liu, Fu, and Zhang}{Geng et~al\mbox{.}}{2023}]%
        {geng2023vip5}
\bibfield{author}{\bibinfo{person}{Shijie Geng}, \bibinfo{person}{Juntao Tan}, \bibinfo{person}{Shuchang Liu}, \bibinfo{person}{Zuohui Fu}, {and} \bibinfo{person}{Yongfeng Zhang}.} \bibinfo{year}{2023}\natexlab{}.
\newblock \showarticletitle{VIP5: Towards Multimodal Foundation Models for Recommendation}.
\newblock \bibinfo{journal}{\emph{arXiv preprint arXiv:2305.14302}}.
\newblock


\bibitem[\protect\citeauthoryear{Hale}{Hale}{2019}]%
        {tubefilter2019}
\bibfield{author}{\bibinfo{person}{James Hale}.} \bibinfo{year}{2019}\natexlab{}.
\newblock \bibinfo{booktitle}{\emph{More Than 500 Hours Of Content Are Now Being Uploaded To YouTube Every Minute}}.
\newblock
\urldef\tempurl%
\url{https://www.tubefilter.com/2019/05/07/number-hours-video-uploaded-to-youtube-per-minute/}
\showURL{%
Retrieved January, 2023 from \tempurl}


\bibitem[\protect\citeauthoryear{Hou, He, McAuley, and Zhao}{Hou et~al\mbox{.}}{2023a}]%
        {hou2023learning}
\bibfield{author}{\bibinfo{person}{Yupeng Hou}, \bibinfo{person}{Zhankui He}, \bibinfo{person}{Julian McAuley}, {and} \bibinfo{person}{Wayne~Xin Zhao}.} \bibinfo{year}{2023}\natexlab{a}.
\newblock \showarticletitle{Learning vector-quantized item representation for transferable sequential recommenders}. In \bibinfo{booktitle}{\emph{TheWebConf}}.
\newblock


\bibitem[\protect\citeauthoryear{Hou, Zhang, Lin, Lu, Xie, McAuley, and Zhao}{Hou et~al\mbox{.}}{2023b}]%
        {hou2023large}
\bibfield{author}{\bibinfo{person}{Yupeng Hou}, \bibinfo{person}{Junjie Zhang}, \bibinfo{person}{Zihan Lin}, \bibinfo{person}{Hongyu Lu}, \bibinfo{person}{Ruobing Xie}, \bibinfo{person}{Julian McAuley}, {and} \bibinfo{person}{Wayne~Xin Zhao}.} \bibinfo{year}{2023}\natexlab{b}.
\newblock \showarticletitle{Large language models are zero-shot rankers for recommender systems}.
\newblock \bibinfo{journal}{\emph{ECIR}}.
\newblock


\bibitem[\protect\citeauthoryear{Ingham}{Ingham}{2023}]%
        {mbw}
\bibfield{author}{\bibinfo{person}{Tim Ingham}.} \bibinfo{year}{2023}\natexlab{}.
\newblock \bibinfo{booktitle}{\emph{Over 60,000 Tracks are Now Uploaded to Spotify Every Day. That's Nearly One per Second.}}
\newblock
\urldef\tempurl%
\url{https://www.musicbusinessworldwide.com/over-60000-tracks-are-now-uploaded-to-spotify-daily-thats-nearly-one-per-second/}
\showURL{%
Retrieved January, 2023 from \tempurl}


\bibitem[\protect\citeauthoryear{Li, Ge, Mao, Li, and Shan}{Li et~al\mbox{.}}{2023a}]%
        {li2023taggpt}
\bibfield{author}{\bibinfo{person}{Chen Li}, \bibinfo{person}{Yixiao Ge}, \bibinfo{person}{Jiayong Mao}, \bibinfo{person}{Dian Li}, {and} \bibinfo{person}{Ying Shan}.} \bibinfo{year}{2023}\natexlab{a}.
\newblock \showarticletitle{TagGPT: Large Language Models are Zero-shot Multimodal Taggers}.
\newblock \bibinfo{journal}{\emph{arXiv preprint arXiv:2304.03022}}.
\newblock


\bibitem[\protect\citeauthoryear{Li, Wang, Li, Fu, Shen, Shang, and McAuley}{Li et~al\mbox{.}}{2023b}]%
        {li2023text}
\bibfield{author}{\bibinfo{person}{Jiacheng Li}, \bibinfo{person}{Ming Wang}, \bibinfo{person}{Jin Li}, \bibinfo{person}{Jinmiao Fu}, \bibinfo{person}{Xin Shen}, \bibinfo{person}{Jingbo Shang}, {and} \bibinfo{person}{Julian McAuley}.} \bibinfo{year}{2023}\natexlab{b}.
\newblock \showarticletitle{Text Is All You Need: Learning Language Representations for Sequential Recommendation}. In \bibinfo{booktitle}{\emph{KDD}}.
\newblock


\bibitem[\protect\citeauthoryear{Li, Zhang, Wang, Xiong, Lu, and Medioni}{Li et~al\mbox{.}}{2023c}]%
        {li2023gpt4rec}
\bibfield{author}{\bibinfo{person}{Jinming Li}, \bibinfo{person}{Wentao Zhang}, \bibinfo{person}{Tian Wang}, \bibinfo{person}{Guanglei Xiong}, \bibinfo{person}{Alan Lu}, {and} \bibinfo{person}{Gerard Medioni}.} \bibinfo{year}{2023}\natexlab{c}.
\newblock \showarticletitle{GPT4Rec: A generative framework for personalized recommendation and user interests interpretation}.
\newblock \bibinfo{journal}{\emph{arXiv preprint arXiv:2304.03879}}.
\newblock


\bibitem[\protect\citeauthoryear{Lin, Shan, Zhu, Du, Chen, Quan, Tang, Yu, and Zhang}{Lin et~al\mbox{.}}{2024}]%
        {lin2023rella}
\bibfield{author}{\bibinfo{person}{Jianghao Lin}, \bibinfo{person}{Rong Shan}, \bibinfo{person}{Chenxu Zhu}, \bibinfo{person}{Kounianhua Du}, \bibinfo{person}{Bo Chen}, \bibinfo{person}{Shigang Quan}, \bibinfo{person}{Ruiming Tang}, \bibinfo{person}{Yong Yu}, {and} \bibinfo{person}{Weinan Zhang}.} \bibinfo{year}{2024}\natexlab{}.
\newblock \showarticletitle{Rella: Retrieval-enhanced large language models for lifelong sequential behavior comprehension in recommendation}.
\newblock \bibinfo{journal}{\emph{TheWebConf}}.
\newblock


\bibitem[\protect\citeauthoryear{Liu, Liu, Lv, Zhou, and Zhang}{Liu et~al\mbox{.}}{2023b}]%
        {liu2023chatgpt}
\bibfield{author}{\bibinfo{person}{Junling Liu}, \bibinfo{person}{Chao Liu}, \bibinfo{person}{Renjie Lv}, \bibinfo{person}{Kang Zhou}, {and} \bibinfo{person}{Yan Zhang}.} \bibinfo{year}{2023}\natexlab{b}.
\newblock \showarticletitle{Is chatgpt a good recommender? a preliminary study}.
\newblock \bibinfo{journal}{\emph{arXiv preprint arXiv:2304.10149}}.
\newblock


\bibitem[\protect\citeauthoryear{Liu, Chen, Sakai, and Wu}{Liu et~al\mbox{.}}{2023a}]%
        {liu2023first}
\bibfield{author}{\bibinfo{person}{Qijiong Liu}, \bibinfo{person}{Nuo Chen}, \bibinfo{person}{Tetsuya Sakai}, {and} \bibinfo{person}{Xiao-Ming Wu}.} \bibinfo{year}{2023}\natexlab{a}.
\newblock \showarticletitle{A First Look at LLM-Powered Generative News Recommendation}.
\newblock \bibinfo{journal}{\emph{arXiv preprint arXiv:2305.06566}}.
\newblock


\bibitem[\protect\citeauthoryear{Mahajan, Porobo~Dharwadker, Shah, Qu, Bang, and Schumitsch}{Mahajan et~al\mbox{.}}{2023}]%
        {mahajan2023pie}
\bibfield{author}{\bibinfo{person}{Khushhall~Chandra Mahajan}, \bibinfo{person}{Amey Porobo~Dharwadker}, \bibinfo{person}{Romil Shah}, \bibinfo{person}{Simeng Qu}, \bibinfo{person}{Gaurav Bang}, {and} \bibinfo{person}{Brad Schumitsch}.} \bibinfo{year}{2023}\natexlab{}.
\newblock \showarticletitle{PIE: Personalized Interest Exploration for Large-Scale Recommender Systems}. In \bibinfo{booktitle}{\emph{Companion Proceedings of the ACM Web Conference 2023}}. \bibinfo{pages}{508--512}.
\newblock


\bibitem[\protect\citeauthoryear{Mansoury, Abdollahpouri, Pechenizkiy, Mobasher, and Burke}{Mansoury et~al\mbox{.}}{2020}]%
        {mansoury2020feedback}
\bibfield{author}{\bibinfo{person}{Masoud Mansoury}, \bibinfo{person}{Himan Abdollahpouri}, \bibinfo{person}{Mykola Pechenizkiy}, \bibinfo{person}{Bamshad Mobasher}, {and} \bibinfo{person}{Robin Burke}.} \bibinfo{year}{2020}\natexlab{}.
\newblock \showarticletitle{Feedback loop and bias amplification in recommender systems}. In \bibinfo{booktitle}{\emph{CIKM}}.
\newblock


\bibitem[\protect\citeauthoryear{Shaw, Uszkoreit, and Vaswani}{Shaw et~al\mbox{.}}{2018}]%
        {shaw2018self}
\bibfield{author}{\bibinfo{person}{Peter Shaw}, \bibinfo{person}{Jakob Uszkoreit}, {and} \bibinfo{person}{Ashish Vaswani}.} \bibinfo{year}{2018}\natexlab{}.
\newblock \showarticletitle{Self-attention with relative position representations}.
\newblock \bibinfo{journal}{\emph{arXiv preprint arXiv:1803.02155}} (\bibinfo{year}{2018}).
\newblock


\bibitem[\protect\citeauthoryear{Song, Sun, Lian, Huang, Li, Jin, and Xie}{Song et~al\mbox{.}}{2022}]%
        {song2022show}
\bibfield{author}{\bibinfo{person}{Yu Song}, \bibinfo{person}{Shuai Sun}, \bibinfo{person}{Jianxun Lian}, \bibinfo{person}{Hong Huang}, \bibinfo{person}{Yu Li}, \bibinfo{person}{Hai Jin}, {and} \bibinfo{person}{Xing Xie}.} \bibinfo{year}{2022}\natexlab{}.
\newblock \showarticletitle{Show me the whole world: Towards entire item space exploration for interactive personalized recommendations}. In \bibinfo{booktitle}{\emph{WSDM}}.
\newblock


\bibitem[\protect\citeauthoryear{Su, Wang, Le, Liu, Li, Lu, Lipshitz, Badam, Heldt, Bi, et~al\mbox{.}}{Su et~al\mbox{.}}{2024}]%
        {su2024long}
\bibfield{author}{\bibinfo{person}{Yi Su}, \bibinfo{person}{Xiangyu Wang}, \bibinfo{person}{Elaine~Ya Le}, \bibinfo{person}{Liang Liu}, \bibinfo{person}{Yuening Li}, \bibinfo{person}{Haokai Lu}, \bibinfo{person}{Benjamin Lipshitz}, \bibinfo{person}{Sriraj Badam}, \bibinfo{person}{Lukasz Heldt}, \bibinfo{person}{Shuchao Bi}, {et~al\mbox{.}}} \bibinfo{year}{2024}\natexlab{}.
\newblock \showarticletitle{Long-Term Value of Exploration: Measurements, Findings and Algorithms}. In \bibinfo{booktitle}{\emph{WSDM}}.
\newblock


\bibitem[\protect\citeauthoryear{Touvron, Martin, Stone, Albert, Almahairi, Babaei, Bashlykov, Batra, Bhargava, Bhosale, et~al\mbox{.}}{Touvron et~al\mbox{.}}{2023}]%
        {touvron2023llama}
\bibfield{author}{\bibinfo{person}{Hugo Touvron}, \bibinfo{person}{Louis Martin}, \bibinfo{person}{Kevin Stone}, \bibinfo{person}{Peter Albert}, \bibinfo{person}{Amjad Almahairi}, \bibinfo{person}{Yasmine Babaei}, \bibinfo{person}{Nikolay Bashlykov}, \bibinfo{person}{Soumya Batra}, \bibinfo{person}{Prajjwal Bhargava}, \bibinfo{person}{Shruti Bhosale}, {et~al\mbox{.}}} \bibinfo{year}{2023}\natexlab{}.
\newblock \showarticletitle{Llama 2: Open foundation and fine-tuned chat models}.
\newblock \bibinfo{journal}{\emph{arXiv preprint arXiv:2307.09288}}.
\newblock


\bibitem[\protect\citeauthoryear{Wang, Lu, Caverlee, Chi, and Chen}{Wang et~al\mbox{.}}{2024}]%
        {wang2024large}
\bibfield{author}{\bibinfo{person}{Jianling Wang}, \bibinfo{person}{Haokai Lu}, \bibinfo{person}{James Caverlee}, \bibinfo{person}{Ed Chi}, {and} \bibinfo{person}{Minmin Chen}.} \bibinfo{year}{2024}\natexlab{}.
\newblock \showarticletitle{Large Language Models as Data Augmenters for Cold-Start Item Recommendation}.
\newblock \bibinfo{journal}{\emph{arXiv preprint arXiv:2402.11724}} (\bibinfo{year}{2024}).
\newblock


\bibitem[\protect\citeauthoryear{Wu, Zhang, Ma, Lyu, He, Mitra, and Liu}{Wu et~al\mbox{.}}{2024}]%
        {wu2024result}
\bibfield{author}{\bibinfo{person}{Haolun Wu}, \bibinfo{person}{Yansen Zhang}, \bibinfo{person}{Chen Ma}, \bibinfo{person}{Fuyuan Lyu}, \bibinfo{person}{Bowei He}, \bibinfo{person}{Bhaskar Mitra}, {and} \bibinfo{person}{Xue Liu}.} \bibinfo{year}{2024}\natexlab{}.
\newblock \showarticletitle{Result Diversification in Search and Recommendation: A Survey}.
\newblock \bibinfo{journal}{\emph{TKDE}} (\bibinfo{year}{2024}).
\newblock


\bibitem[\protect\citeauthoryear{Xi, Liu, Lin, Zhu, Chen, Tang, Zhang, Zhang, and Yu}{Xi et~al\mbox{.}}{2023}]%
        {xi2023towards}
\bibfield{author}{\bibinfo{person}{Yunjia Xi}, \bibinfo{person}{Weiwen Liu}, \bibinfo{person}{Jianghao Lin}, \bibinfo{person}{Jieming Zhu}, \bibinfo{person}{Bo Chen}, \bibinfo{person}{Ruiming Tang}, \bibinfo{person}{Weinan Zhang}, \bibinfo{person}{Rui Zhang}, {and} \bibinfo{person}{Yong Yu}.} \bibinfo{year}{2023}\natexlab{}.
\newblock \showarticletitle{Towards Open-World Recommendation with Knowledge Augmentation from Large Language Models}.
\newblock \bibinfo{journal}{\emph{arXiv preprint arXiv:2306.10933}}.
\newblock


\bibitem[\protect\citeauthoryear{Yang, Yi, Zhiyuan~Cheng, Hong, Li, Xiaoming~Wang, Xu, and Chi}{Yang et~al\mbox{.}}{2020}]%
        {yang2020mixed}
\bibfield{author}{\bibinfo{person}{Ji Yang}, \bibinfo{person}{Xinyang Yi}, \bibinfo{person}{Derek Zhiyuan~Cheng}, \bibinfo{person}{Lichan Hong}, \bibinfo{person}{Yang Li}, \bibinfo{person}{Simon Xiaoming~Wang}, \bibinfo{person}{Taibai Xu}, {and} \bibinfo{person}{Ed~H Chi}.} \bibinfo{year}{2020}\natexlab{}.
\newblock \showarticletitle{Mixed negative sampling for learning two-tower neural networks in recommendations}. In \bibinfo{booktitle}{\emph{Companion Proceedings of the Web Conference 2020}}.
\newblock


\bibitem[\protect\citeauthoryear{Yu, Wu, Wu, Yi, and Liu}{Yu et~al\mbox{.}}{2022}]%
        {yu2021tiny}
\bibfield{author}{\bibinfo{person}{Yang Yu}, \bibinfo{person}{Fangzhao Wu}, \bibinfo{person}{Chuhan Wu}, \bibinfo{person}{Jingwei Yi}, {and} \bibinfo{person}{Qi Liu}.} \bibinfo{year}{2022}\natexlab{}.
\newblock \showarticletitle{Tiny-newsrec: Effective and efficient plm-based news recommendation}. In \bibinfo{booktitle}{\emph{EMNLP}}.
\newblock


\end{thebibliography}

% \subsection{Offline Evaluation}
% \label{sec:offline}
% \textbf{might move to Appendix}
% During the offline development phase, we also set aside a test set consisting of 10K $[(C_1, C_2), C_L]$ examples. We examine the probability that the next “new” cluster inferred by the fine-tuned LLM can hit the ground-truth, to get a sense of how well the LLM is aligned with user behavior. We include these baseline in the comparison: a) we randomly select one candidate cluster as prediction (no LLM); b) use LLM finetuned with ~8k randomly-generated [(cluster 1, cluster 2), label] samples; c) LLM finetuned with non-diversify dataset.

% \begin{table}[h]
% \caption{Offline recall for different variations.}
% \resizebox{0.3\textwidth}{!}{%
% \begin{tabular}{l|llll}
% \hline
% Model & a & b & c & d    \\ \hline \hline
% Hit Rate & 0.005 & 0.004 & 0.011 & 0.018\\ \hline
% \end{tabular}%
% }
% \label{tab:offline_recall}
% \end{table}

% Though b) is able to generate clusters which can be perfectly mapped to existing clusters (i.e., great formatting capability), it obtains much lower recall compared to c) and d), and performs similar to a). It demonstrates that our finetuned model not only obtained formatting capability but also successfully got alignment with YouTube user behavior with the finetuning process. And the improvement from c) to d) also showcases the effectiveness of our data curation process.

\end{document}